\documentclass[8.5pt,twoside,twocolumn]{article}
\usepackage[super,sort&compress,comma]{natbib} 
\usepackage[version=3]{mhchem}
\usepackage{times,mathptmx}
\usepackage{sectsty}
\usepackage{balance} 
\usepackage[english]{babel}

\usepackage{color}
\usepackage{units}
\usepackage{graphicx} 
\usepackage{lastpage}
\usepackage[format=plain,singlelinecheck=false,font=small,labelfont=bf,labelsep=space]{caption} 
\usepackage{fancyhdr}
\begin{document}

\thispagestyle{plain}
\fancypagestyle{plain}{}
\renewcommand{\headrulewidth}{1pt}
\renewcommand{\thefootnote}{\fnsymbol{footnote}}
\renewcommand\footnoterule{\vspace*{1pt}%
\hrule width 3.4in height 0.4pt \vspace*{5pt}} 
\setcounter{secnumdepth}{5}

\makeatletter 
\def\subsubsection{\@startsection{subsubsection}{3}{10pt}{-1.25ex plus -1ex minus -.1ex}{0ex plus 0ex}{\normalsize\bf}} 
\def\paragraph{\@startsection{paragraph}{4}{10pt}{-1.25ex plus -1ex minus -.1ex}{0ex plus 0ex}{\normalsize\textit}} 
\renewcommand\@biblabel[1]{#1}            
\renewcommand\@makefntext[1]%
{\noindent\makebox[0pt][r]{\@thefnmark\,}#1}
\makeatother 
\renewcommand{\figurename}{\small{Fig.}~}
\sectionfont{\large}
\subsectionfont{\normalsize} 

\fancyfoot{}
\fancyfoot[RO]{\footnotesize{\sffamily{1--\pageref{LastPage} ~\textbar  \hspace{2pt}\thepage}}}
\fancyfoot[LE]{\footnotesize{\sffamily{\thepage~\textbar\hspace{3.45cm} 1--\pageref{LastPage}}}}
\fancyhead{}
\renewcommand{\headrulewidth}{1pt} 
\renewcommand{\footrulewidth}{1pt}
\setlength{\arrayrulewidth}{1pt}
\setlength{\columnsep}{6.5mm}
\setlength\bibsep{1pt}

\twocolumn[
  \begin{@twocolumnfalse}
\noindent\LARGE{\textbf{Exposing nanobubble-like objects to a degassed environment}}
\vspace{0.6cm}

\noindent\large{\textbf{Robin P. Berkelaar,$^{\ast}$\textit{$^{a,b,c}$} Erik Dietrich,\textit{$^{b,c}$} Gerard A. M. Kip,\textit{$^{d}$} E. Stefan Kooij,\textit{$^{b}$} Harold J. W. Zandvliet,$^{\ast}$\textit{$^{b}$}and
Detlef Lohse$^{\ast}$\textit{$^{c}$}}}\vspace{0.5cm}

\noindent\textit{\small{\textbf{Received Xth XXXXXXXXXX 20XX, Accepted Xth XXXXXXXXX 20XX\newline
First published on the web Xth XXXXXXXXXX 200X}}}

\noindent \textbf{\small{DOI: 10.1039/b000000x}}
\vspace{0.6cm}

\noindent \normalsize{The primary attribute of interest of surface nanobubbles is their unusual stability and a number of theories trying to explain this have been put forward. Interestingly, the dissolution of nanobubbles is a topic that did not receive a lot of attention yet. In this work we applied two different experimental procedures which should cause gaseous nanobubbles to completely dissolve. In our experiments we nucleated nanobubble-like objects by putting a drop of water on HOPG using a plastic syringe and disposable needle. In method A, the nanobubble-like objects were exposed to a flow of degassed water ($1.17~\unit{mg/l}$) for $96~\unit{hours}$. In method B, the ambient pressure was lowered in order to degas the liquid and the nanobubble-like objects. Interestingly, the nanobubble-like objects remained stable after exposure to both methods. After thorough investigation of the procedures and materials used during our experiments, we found that the nanobubble-like object were induced by the use of disposable needles in which PDMS contaminated the water. It is very important for the nanobubble community to be aware of the fact that, although features look and behave like nanobubbles, in some cases they might in fact be or induced by contamination. The presence of contamination could also resolve some inconsistencies found in the nanobubble literature.}
\vspace{0.5cm}
 \end{@twocolumnfalse}
  ]

\section{Introduction}


\footnotetext{\textit{$^{a}$~Materials innovation institute (M2i), 2628 CD Delft, The Netherlands, E-mail: r.p.berkelaar@utwente.nl}}
\footnotetext{\textit{$^{b}$~Physics of Interfaces and Nanomaterials, MESA+ Institute for Nanotechnology, University of Twente, P.O. Box 217, 7500 AE Enschede, The Netherlands, E-mail: h.j.w.zandvliet@utwente.nl }}
\footnotetext{\textit{$^{c}$~Physics of Fluids and J.M. Burgers Centre for Fluid Mechanics, MESA+ Institute for Nanotechnology, University of Twente, P.O. Box 217, 7500 AE Enschede, The Netherlands, E-mail: d.lohse@utwente.nl }}
\footnotetext{\textit{$^{d}$~MESA+ Institute for Nanotechnology, University of Twente, P.O. Box 217, 7500 AE Enschede, The Netherlands }}



Sub-micron bubbles on hydrophobic interfaces were thought to be the cause of the long-range hydrophobic interaction and this resulted in the emergence of a completely new field, that of surface nanobubbles~\cite{hampton2010,craig2010,seddon2011}. The seminal work was performed by Parker and Attard~\cite{parker1994} in 1994, who observed long-range attractive forces using a surface force apparatus and attributed this to the existence of nano-scale gas bubbles at the interface. Real-space images of nanobubbles had to wait until advancements in atomic force microscopy (AFM) immersed in liquids resulted in the observation of soft spherical cap shaped features by Lou et al.~\cite{lou2000} and Ishida et al.~\cite{ishida2000} in 2000. Unfortunately, the AFM tip disturbs these soft features and properly imaging nanobubbles is not a trivial task~\cite{walczyk2013,wang2013a,zhao2013,borkent2010}. Nanobubbles have been observed on a wide variety of surfaces\cite{wang2013,holmberg2003,agrawal2005,yang2007,wang2009,zhang2011} and found to be stable in a broad range of conditions like elevated temperature~\cite{yang2007,guan2012,berkelaar2012}, low pH~\cite{zhang2006c} and salt solutions~\cite{mazumder2011,zhang2006c}. 

The first convincing proof for the gaseous nature of these features came from Zhang et al.~\cite{zhang2007,zhang2008}, showing gas-enrichment near the interface using ATR-IR measurements in 2007. During this period, the gaseous nature of these bubbles was also indirectly inferred by degassing the liquids used to nucleate nanobubbles~\cite{zhang2004,zhang2006d} and by degassing the nanobubble covered substrate~\cite{zhang2006}. Also the growth of nanobubbles by rectified diffusion using an acoustic field suggests their gaseous nature~\cite{brotchie2011}. Although an abundance of experiments have been performed on these nanobubbles since 1994, mixed results on a number of topics were found.  For example, the contact angle depends in some studies on the radius of curvature~\cite{yang2003,limbeek2011}, whereas in other experiments the contact angle is found to be constant~\cite{zhang2006c,borkent2010}. The presence of a gas layer at the solid-liquid interface is observed in several experimental studies~\cite{miller1999,steitz2003}, where in other studies such a gaseous phase is not found~\cite{poynor2006,mezger2006}. Also, nanobubbles are sometimes found in ethanol~\cite{simonsen2004,ishida2000}, while others observe pristine surfaces if immersed in ethanol~\cite{zhang2004}. And finally, nanobubbles are in some cases imaged on HOPG just by immersing the substrate in water~\cite{walczyk2013,borkent2007,berkelaar2013} while others need to perform an ethanol-water exchange to induce nanobubble nucleation~\cite{zhang2011a}.

Despite the inconsistencies, what the nanobubbles all have in common is their long term stability. The fact that these bubbles can be observed is quite a remarkable feat on its own, they have  been measured to be stable for as long as several days~\cite{zhang2008,zhang2013}. For small bubbles the Laplace pressure dominates, and this drives the dissolution of gas from the bubble into the liquid. Bubbles with a radius of curvature $R_c$ less than $1~\unit{\mu m}$ should thus dissolve on a timescale of $\tau \sim R_{c}^2/D$, where $D\approx 1 \cdot 10^{-9} ~m^{2}/ s$, i.e. in microseconds~\cite{epstein1950,ljunggren1997}. The existence of stable bubbles with radii of a few hundred nanometers and heights in the order of ten nanometers, hence the name nanobubbles, sparked the interest to what the mechanism behind this remarkable stability could be.

Since the discovery of nanobubbles a number of theories explaining this surprising behavior have been proposed. Just after the discovery of nanobubbles it was argued they might not be bubbles, but contamination, e.g. resulting from polymers used to hydrophobize the surface (theory 1)~\cite{evans2004}. However, this was soon to be discarded by the assertion that the bubble contained gas. A new theory followed, in which the presence of contamination at the bubble gas-liquid interface lowered the surface tension, and thus lowered the Laplace pressure, which in effect reduced the dissolution of the bubble (theory 2)~\cite{ducker2009}. In addition, calculations of the contamination concentration needed for a sufficiently low surface tension to match the measured contact angle for nanobubbles, resulted in a layer thickness which greatly hinders the gas out-flux~\cite{ducker2009}. Also the calculations from Das et al.~\cite{das2010} suggests that a possible contamination lowers the surface tension and the gas-flux through the interface, but this was insufficient to stabilize the bubble.
Experiments using a surfactant to remove a hypothetical contamination layer by Zhang et al.~\cite{zhang2006c} showed that nanobubbles remain stable and do not dissolve when exposed to SDS surfactant (which should wash away contaminations), a result confirmed by Peng et al.~\cite{peng2013}. As these authors showed, the detergents though help to mechanically remove surface nanobubbles with the AFM tip.  Both used a surfactant concentration below the critical micelle concentration (CMC).  In other studies a concentration above the CMC was used; Ducker~\cite{ducker2009} showed the dissolution of nanobubbles in this particular case. However, in more recent work Zhang et al.~\cite{zhang2012} observed again stable nanobubbles even for surfactant concentrations above the CMC. 

As the stability could not be explained by contamination, there was need for a new and completely different approach, which resulted in the dynamic equilibrium theory by Brenner and Lohse~\cite{brenner2008} (theory 3). The main idea of this theory is that the gas out-flux of the bubble is compensated by a gas in-flux at the three-phase contact line. This theory was later extended and specified by Seddon et al.~\cite{seddon2011c}: the gas inside a nanobubble fulfills the requirements for a Knudsen gas, meaning that the mean free path of the gas molecules is larger than the distance to the interface of the bubble. Therefore, gas-molecules desorbing from the gas-solid interface will hit the gas-liquid interface and transfer momentum along a preferred direction perpendicular to the solid-liquid interface. This then drives a circulatory flow around the nanobubble transporting a stream of gas rich water to the three-phase line of the nanobubble, where the gas adsorbs onto the surface and diffuses back into the nanobubble. Using alternate formulations of this theory made it possible to predict the temperature and gas saturation dependency of nanobubbles~\cite{petsev2013}. What, however, remains unclear in this theory is what energetically drives the flow and therefore a non-equilibrium situation has to be assumed. 

Recently another theory was proposed by Weijs and Lohse~\cite{weijs2013} (theory 4), which does not suffer from the difficulty that the dynamic equilibrium theory has. The theory combines the assumption, which is in some cases observed experimentally~\cite{zhang2013}, that the contact line of a nanobubble is pinned together with the retardation of gas diffusion in a liquid compared to air. The moment a small amount of gas leaves the nanobubble, the contact angle will have to decrease in order to accommodate the reduction in volume, which in return lowers the Laplace pressure and hence slows down the dissolution of the nanobubble. The gas molecules, which just dissolved from the bubble into the liquid, increase the gas saturation around the nanobubble and will take a significant time to diffuse to the interface of the water film and leave the system. The increased gas saturation around the nanobubble, resulting from these gas molecules and those from neighboring nanobubbles, lowers the out-flux of new gas molecules from the nanobubble and thus enhances the stabilization. The combined effect of contact-line pinning and diffusion retardation in liquids results in considerable longer lifetimes, dependent on the liquid-film thickness $\tau \sim \ell^2/D$, where $\ell$ is the liquid film thickness and D the diffusion coefficient of gas in liquid. 

The number of experimental studies that focused on verifying or disproving one or more of the above theories is rather limited and the results from these studies are often inconsistent. 
A recent experiment showed that nanobubbles were stable in degassed water which was refreshed every 20 minutes, and the authors concluded that this was most likely due to contamination~\cite{wang2013}. However, Zhang et al.~\cite{zhang2006} have shown the localized disappearance of nanobubbles after degassing. though some regions remain covered with nanobubbles also after degassing.

The scope of this present study is to try  to contribute to a clarification of the puzzling situation. In the way to produce nanobubbles or nanobubble-like objects, we will restrict us to the case of liquid deposition on hydrophobic flat surfaces. We will not address the most popular method for nanobubble formation, namely ethanol-water exchange, or more generally, solvent-exchange. We investigate whether the nanobubble-like objects "communicate" with the surrounding liquid by the diffusion of gas molecules using two different methods. In method A, the nanobubble-like objects are exposed to a continuous flow of degassed water and in method B the ambient pressure is reduced. This should unambiguously result in a significant reduction of the lifetime, if nanobubbles are stabilized by theory 3, theory 4 or any other mechanism where the gas can diffuse through the gas-liquid interface of the bubble. In both cases the nanobubble-like objects, which develop at deposition, remain stable after a prolonged exposure to degassed water. We conclude that therefore they are either actually not gaseous or have a gas impermeable shell, which could be in accordance with theories 1 or 2, i.e. due to contamination, or the result of an, so far, unknown physical mechanism.

\section{Experimental details}

Nucleation of nanobubble-like objects was acquired by immersing an HOPG (ZYA grade, MikroMasch) substrate in water. The substrate was freshly cleaved prior to each experiment and subsequently clamped between two Teflon rods in an all Teflon liquid-cell. The liquid-cell was cleaned in a Piranha solution (a 3:1 H$_{2}$SO$_{4}$ to 30\% H$_{2}$O$_{2}$ mixture) and rinsed with copious amounts of water. Purification of the water was performed by a Simplicity 185 system (Millipore) up to a resistivity of $18.2~\unit{M\Omega \cdot cm}$. The liquid-cell was filled with $3-4~\unit{ml}$ water using a new $5~\unit{ml}$ sterile plastic disposable syringe (Discardit, BD) and disposable needle (Microlance, BD). The Teflon liquid cell was then placed within an Agilent 5100 atomic force microscope. The AFM nose-cone was rinsed thoroughly with ethanol (Emsure $\geq$ 99.9\% purity, Merck) and dried in a N$_{2}$ gas flow before imaging. 
The immersed HOPG surface covered with nanobubbles was imaged by the AFM operated in intermittent contact mode. The liquid-cell was subsequently removed from the AFM and sealed with a SiO$_{2}$ wafer, which was cleaned in Piranha solution and rinsed with water. Thereafter, the liquid cell was purged with degassed water up to $96~\unit{hours}$. Finally, the effect of degassed water flow was checked by a renewed scan of the identical position on the HOPG surface.

Degassing was performed in a glass vessel, filled with $1.4~\unit{L}$ water, by reducing the pressure to $P_{e} \approx 20~\unit{mbar}$ using a membrane pump (MD-4T, Vacuubrand). The water was stirred and temperature controlled at $21^\circ ~\unit{C}$ (RCT basic \& ETS-D4, IKA Werke) while degassing. The steady state oxygen saturation inside the glass vessel was measured (Presens, recently calibrated) to be $<4\%$ ($0.36~\unit{mg/L}$). The degassed water was extracted from the glass vessel through Teflon tubing and a small piece of flexible R3603 Tygon tube using a peristaltic pump (Model 7519-05, Masterflex). The glass vessel was continuously pumped, to ensure a low gas concentration throughout the experiment, while extracting degassed water at a rate of $1.5~\unit{ml/min}$. Water in an identical secondary set-up was degassed in parallel and the extraction of degassed water was switched between set-ups when the water level in one of them became low. Switching between the two set-ups was performed within $10~\unit{s}$ and this procedure guaranteed a continuous flow of degassed water up to the maximum experiment duration of $96~\unit{hours}$. Measurement of the steady state O$_{2}$ gas saturation inside the liquid-cell, during degassed flow, was $<13\%$ ($1.17~\unit{mg/l}$). The O$_{2}$ gas saturation dropped at the start of the experiment towards the steady state value of $<13\%$ with a time constant of $\tau=1.3~\unit{h}$.
Imaging was performed using Al-back-coated NSC36c Si$_{3}$N$_{4}$ probes obtained from MikroMasch, with a nominal spring-constant of $0.6~\unit{N/m}$, resonance frequency of $\omega_{0}= 65~\unit{kHz}$ (dry environment), resonance frequency in water of $\omega_{0,w}= 34~\unit{kHz}$, and tip radius of $8~\unit{nm}$. The set-point was kept as high as possible ($\sim$95\%) and the amplitude was chosen in the range of  20-30 nm, in order to minimize the deformation of the nanobubbles by the tip. 

For the X-ray photoelectron spectroscopy (XPS) measurements a Quantera SXM (Physical Electronics) was used. The X-rays were Al K$\alpha$, monochromatic at 1486.6 eV with a beam size of 200~$\mu m$. On every sample 4 different areas were probed with an area size of 600 $\times$ 300 $\mu m^2$.

\section{Results \& Discussion}

\begin{figure}[h]
\centering
\includegraphics[width=1\columnwidth]{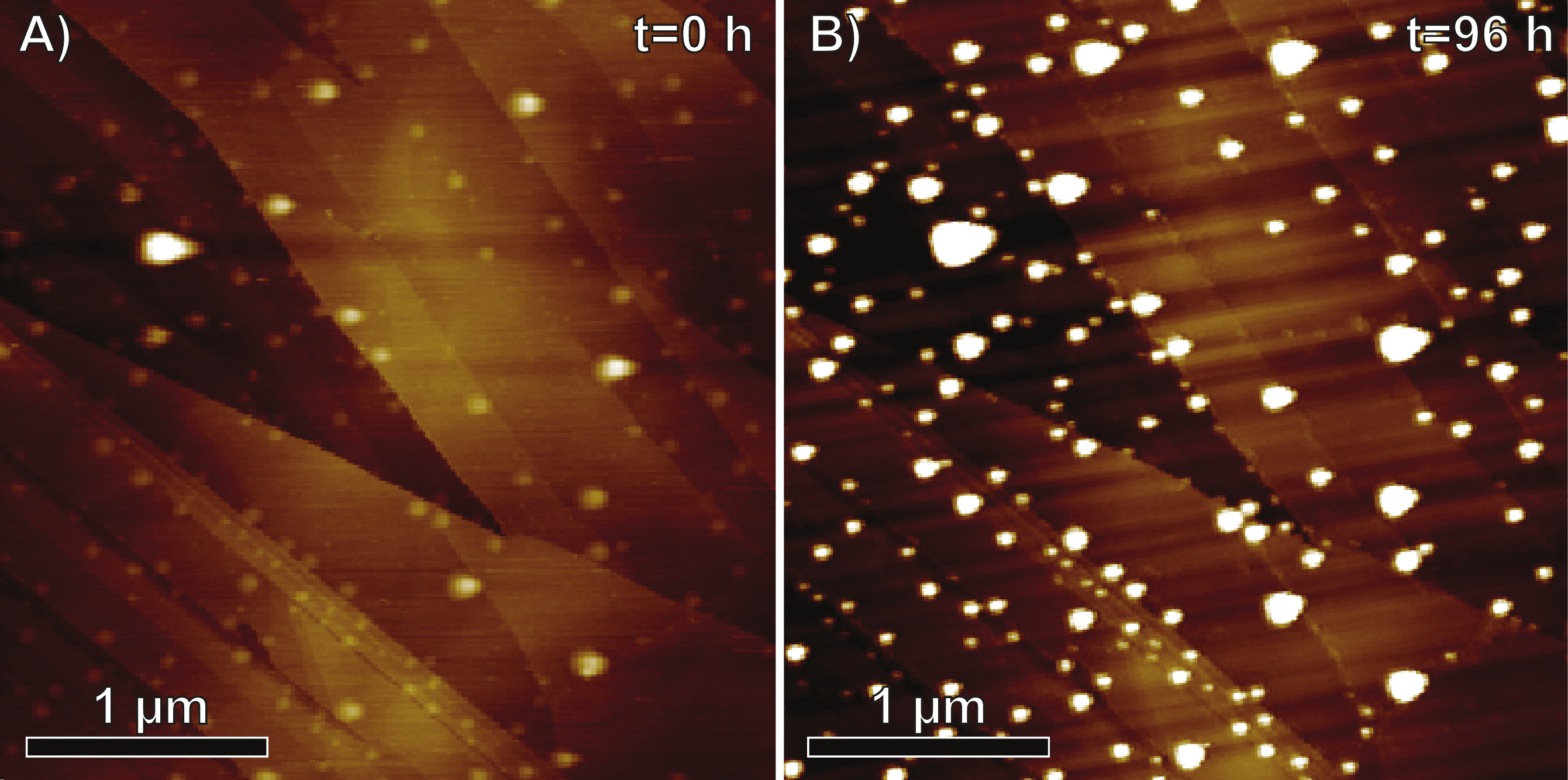}

\caption[]{ (color online) AFM images of an HOPG surface immersed in water. Nanobubble-like objects (appearing as bright features) were nucleated by applying a droplet of water on dry HOPG, using a plastic syringe and disposable needle (A). After flowing degassed water over the surface for 96 hours the objects still remain (B).  The measured O$_{2}$ gas saturation in the liquid cell during the flow was  $<15\%$. The $z$-range is $14~\unit{nm}$. }
\label{fig:flow}
\end{figure}

The stability of our nanobubble-like objects was first challenged using method A: degassed water was flowed over the objects for a prolonged time. A freshly cleaved HOPG surface was clamped into an all Teflon liquid cell and immersed in water. The liquid-cell was then mounted into the AFM, where the surface was scanned in intermittent-contact mode. This procedure resulted in a substantial coverage of surface nanobubble-like objects, as can be observed in Figure~\ref{fig:flow}A. The larger objects have an asymmetrical appearance, generally referred to as parachuting, due to the set-point being close to 100\%.  The set-point was intentionally adjusted close to 100\% in order to limit the deformation of the objects by the tip. The liquid-cell was thereafter removed from the AFM and closed using a SiO$_{2}$ substrate. Degassed water was then injected into the liquid cell with a continuous flow of $1.5~\unit{ml/min}$. As a result, the measured O$_{2}$ saturation inside the liquid cell during degassed water flow was $<13\%$ ($1.17~\unit{mg/l}$). The diffusion coefficients at $20^{\circ}$ C in water for the other major constituents of air (nitrogen $D=2.6 \cdot 10^{-9}~\unit{m^{2}/s}$ and argon $D=2.3 \cdot 10^{-9}~\unit{m^{2}/s}$) are comparable to that of oxygen ($D=2.3 \cdot 10^{-9}~\unit{m^{2}/s}$)~\cite{difgas}. The measured O$_{2}$ saturation can thus be regarded as the absolute gas saturation of the water. Flowing degassed water has a number of advantages compared to other degassing techniques. Firstly, a continuously low gas-saturation can be guaranteed, even when the liquid-cell is not sealed properly. Secondly, the flow will cause convection and thus better mixing compared to statically filling the liquid cell with degassed water. Thirdly, there are no detrimental effects from macroscopic bubbles expanding and sweeping clean the area of interest, as can be the case for degassing by reducing the ambient pressure~\cite{zhang2006}. After exposing the nanobubble-like objects to the degassed water flow for $96~\unit{hours}$ the liquid-cell was placed back in the AFM, and the same area was imaged once more.

Surprisingly, the nanobubble-like features had not vanished, quite the opposite, they appear even larger, Figure~\ref{fig:flow}B. All effort was taken to exclude deformation of the objects by the tip and having similar scanning parameters for all images, such as amplitude ($\unit{nm}$) and set-point. Still we have the impression that a parameter has changed, such as liquid height or the difference in effective spring-constant between the cantilevers used in the two images (though freshly taken from the same box), and this increase does not represent an actual change in the objects' height. In any case, we are not so much concerned about the actual geometry of the nanobubble-like objects, rather the fact that they are still present after prolonged exposure to a degassed environment is of interest. Comparing the images before and after degassed water flow, results in virtually no change regarding the number and position of the nanobubble-like objects. The only viable explanation is the absence of mass transfer through the interface between the nanobubble-like objects and the water.

%
%

\begin{figure}[h]
\centering
\includegraphics[width=1\columnwidth]{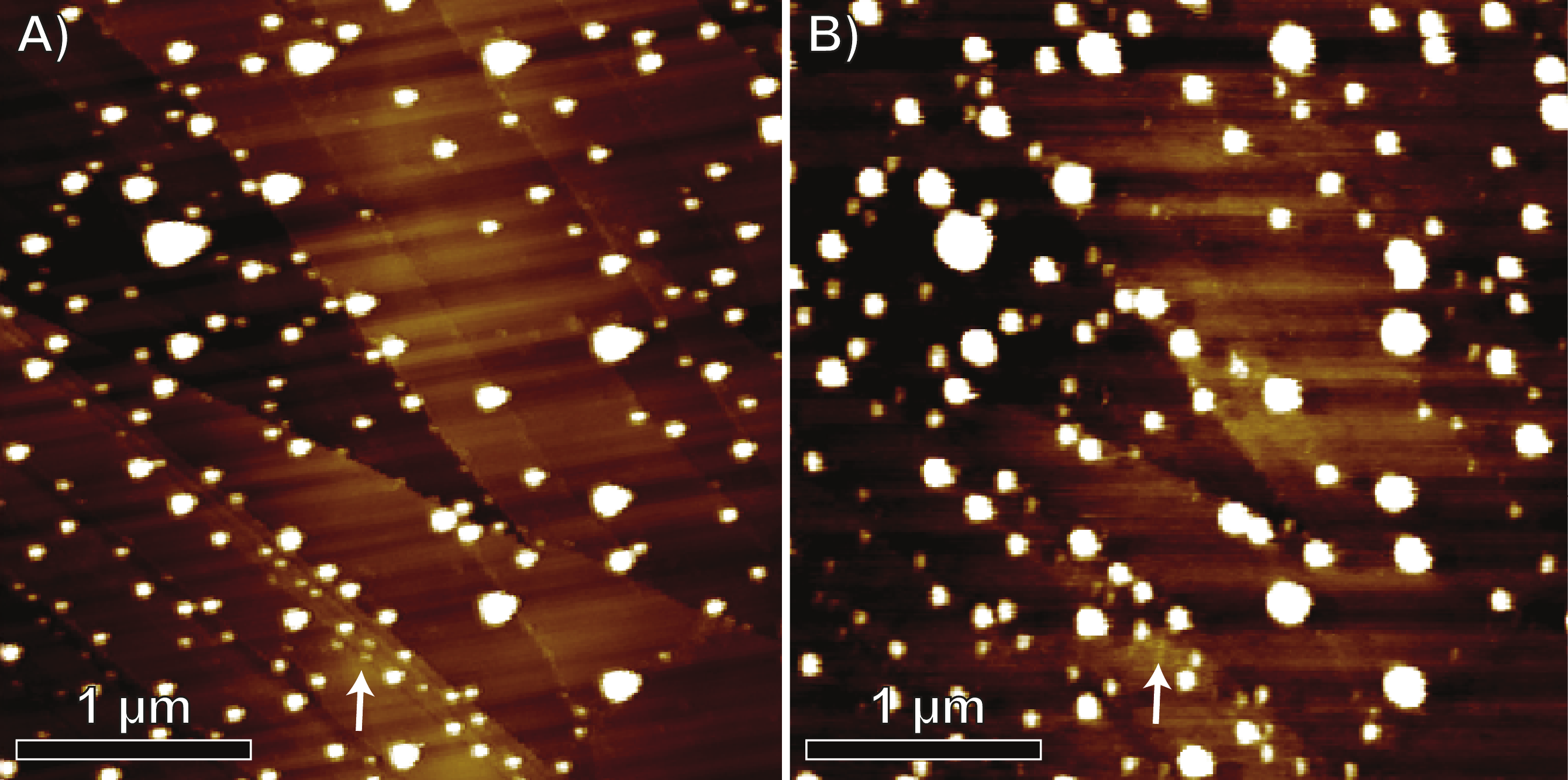}
\caption{ (color online) AFM images of an HOPG surface immersed in Millipore  water. The bright features observed on the surface are nanobubble-like objects nucleated by depositing a drop of water on HOPG, after being exposed to degassed water (A). The drop deposition was done using a plastic syringe and disposable needle. The sample was transferred to a pressure chamber after it was imaged. The pressure was subsequently lowered from atmospheric pressure to $\approx 20~\unit{mbar}$ during a period of $24~\unit{h}$, followed by $30~\unit{min}$ degassing at a stable pressure of $\approx 20~\unit{mbar}$. The pressure chamber was pressurized to atmospheric pressure and the sample was transferred back into the AFM. The number and position of the nanobubble-like objects show little change after the degassing procedure (B). The arrows point to small nanobubble-like objects which are not visible after degassing, presumably due to the reduced resolution. The $z$-range is $14~\unit{nm}$. }
\label{fig:pump}
\end{figure}

We also investigated the gas out-flux using the above mentioned method B, in which the nanobubble-like objects are exposed to a reduced pressure for a prolonged period of time. The  liquid cell with the immersed HOPG sample was removed from the AFM and inserted into a glass pressure vessel. A few centimeters high water layer was present in the pressure vessel to prevent complete evaporation of the water in the liquid cell. The pressure was gradually dropped from atmospheric pressure to $\approx 20~\unit{mbar}$ in the course of $24~\unit{h}$. It is essential that the pressure drop is slow, in order to restrain the formation of macroscopic bubbles on the HOPG interface, since the moving contact line of a growing macroscopic bubble will efface the nanobubble-like objects from the surface. The pressure remained at a low pressure of $\approx 20~\unit{mbar}$ for $30~\unit{min}$ before increasing it back to atmospheric conditions within $5~\unit{min}$. The liquid-cell was subsequently returned to the AFM for imaging. The results are similar to that of the degassed water flow experiment. Comparing the same area before and after degassing reveals that again the number and size of the nanobubble-like objects remains virtually unchanged, see Figure~\ref{fig:pump}. Some of the smallest objects do not appear on the image after degassing (an example is pointed out using arrows), Figure~\ref{fig:pump}B, which is presumably due to  reduced resolution. This is in agreement with the results from the degassed water flow, there does not seem to be any mass-transfer between the alleged nanobubble-like objects and the liquid.

%
%

A bubble with a gas-impermeable shell should still expand in volume, which could result in changes in the nanobubble coverage as discussed in Appendix A. However, there we show that the changes in radius or contact angle are too small to result in any lasting modifications by coalescence in nanobubble coverage after the pressure is reduced and subsequently increased back to atmospheric conditions. 

The results from the depressurizing experiment are in complete agreement with the degassed water flow experiment: In both cases the nanobubble-like objects do not dissolve. This again implies that there is no mass transfer between the nanobubble-like object and the liquid. Two stability theories (3 \& 4), the dynamic equilibrium theory and limited diffusion theory, both depend on a mechanism that involves gas in- and outflux. Therefore these two theories are in contradiction to the present results for the analyzed features and, for these nanobubble-like objects created by deposition, we have to turn our attention to the two remaining theories. Either these nanobubble-like objects have a gas-impermeable shell or these objects are simply not bubbles, but droplets of contamination.

Both theories depend on a certain concentration of contaminants present in the system. Investigating the literature revealed that a variety of contamination sources could possibly play a role. These sources include, but are not limited to: glue from the adhesive tape used for cleaving HOPG~\cite{rabe2013}, poor quality solvents~\cite{habich2010}, plastic syringes~\cite{syringe1,syringe2}, flexible tubes, and air quality. Also, when employing the ethanol-exchange to nucleate nanobubbles a lot of care has to be taken as ethanol is especially susceptible to distribute any organic contaminants present in the nucleation procedure. 

\begin{figure}[h]
\centering
\includegraphics[width=1\columnwidth]{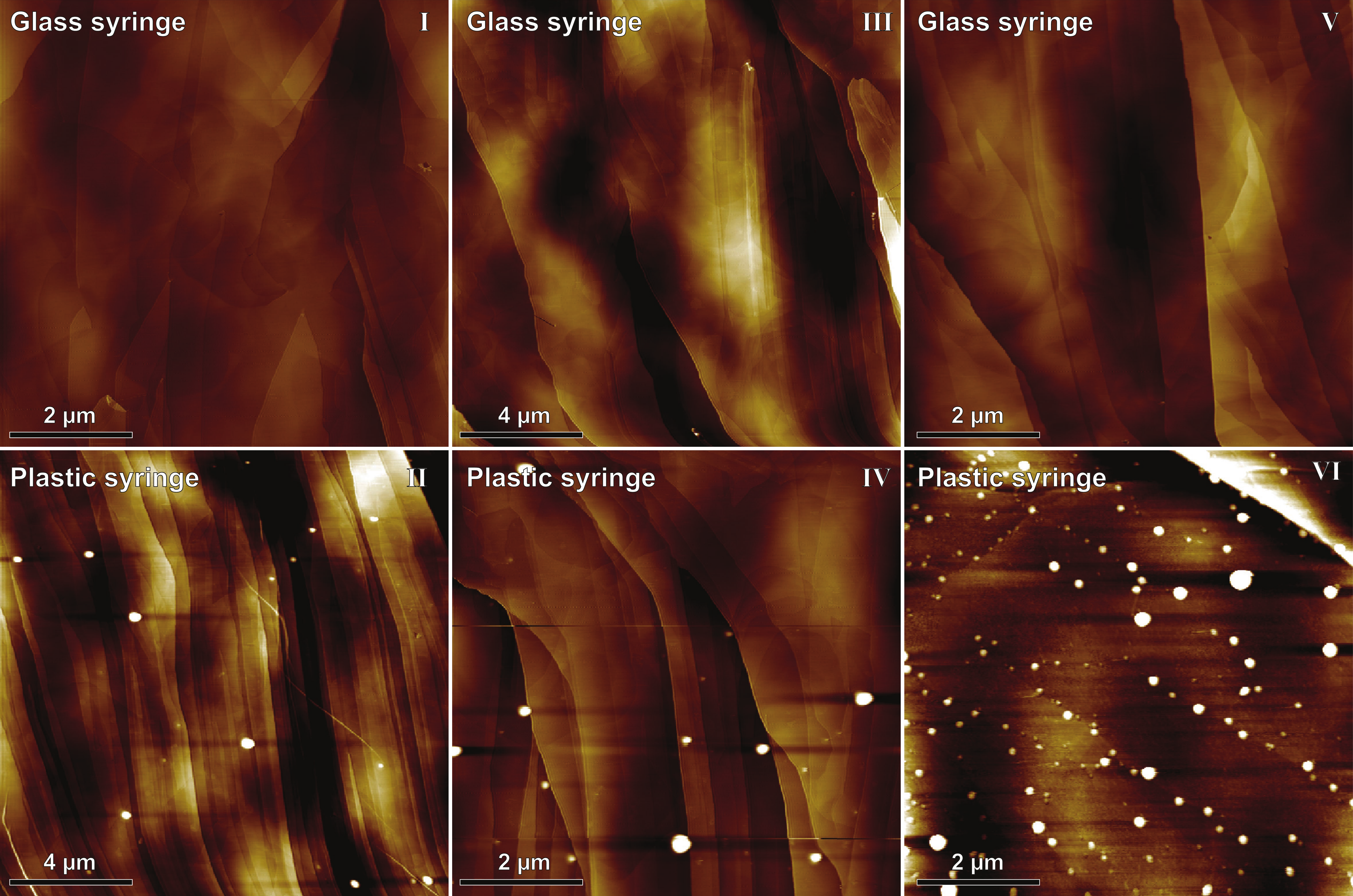}
\caption{ (color online) AFM images of an HOPG surface under a droplet of water for six different experiments performed in the order I-VI. The droplet of water was deposited to the freshly cleaved HOPG surface using a glass syringe and full-metal needle (in experiments I, III, V) or a plastic syringe and disposable needle (in experiments II, IV, VI). Only when using the plastic syringes with disposable needles nanobubble-like objects are observed. The $z$-range is $18 ~\unit{nm}$. }
\label{fig:syringe}
\end{figure}

The procedures and materials used in our deposition experiment were scrutinized for any possible contaminant sources. Nonetheless, in our case it turned out that the use of sterile disposable plastic syringes and/or disposable needles was a crucial step for the nucleation of the nanobubble-like objects. We checked this finding by depositing a drop of water on freshly cleaved HOPG using either a glass syringe and full-metal needle or a questionable plastic syringe and disposable needle. Figure~\ref{fig:syringe} shows six experiments, labeled and performed in the order I-VI, in which a glass (in experiments I, III, V) or disposable plastic syringe and disposable needle (in experiments II, IV, VI) were used to deposit the water on the substrate. It clearly shows that no nanobubble-like objects were observed if a glass syringe was used, however, in the case of a plastic syringe with disposable needle objects looking like nanobubbles are found. In both cases different positions on the HOPG sample were imaged with similar results. The water was kept in plastic syringes for durations ranging from a few minutes up to a day, which resulted in no significant changes in the nanobubble-like coverage. However, refilling the plastic syringe with water several times does result in a reduced surface coverage with nanobubble-like objects. These degassing results are different from what was observed by Zhang et al. after degassing, where they show regions on the HOPG substrate where nanobubbles have disappeared~\cite{zhang2006}. This can be explained by having used a procedure that does not introduce contamination and produces gaseous nanobubbles. However, this does not explain why nanobubbles remained stable in other regions.

%
%

\begin{figure}
\centering
\includegraphics[width=1\columnwidth]{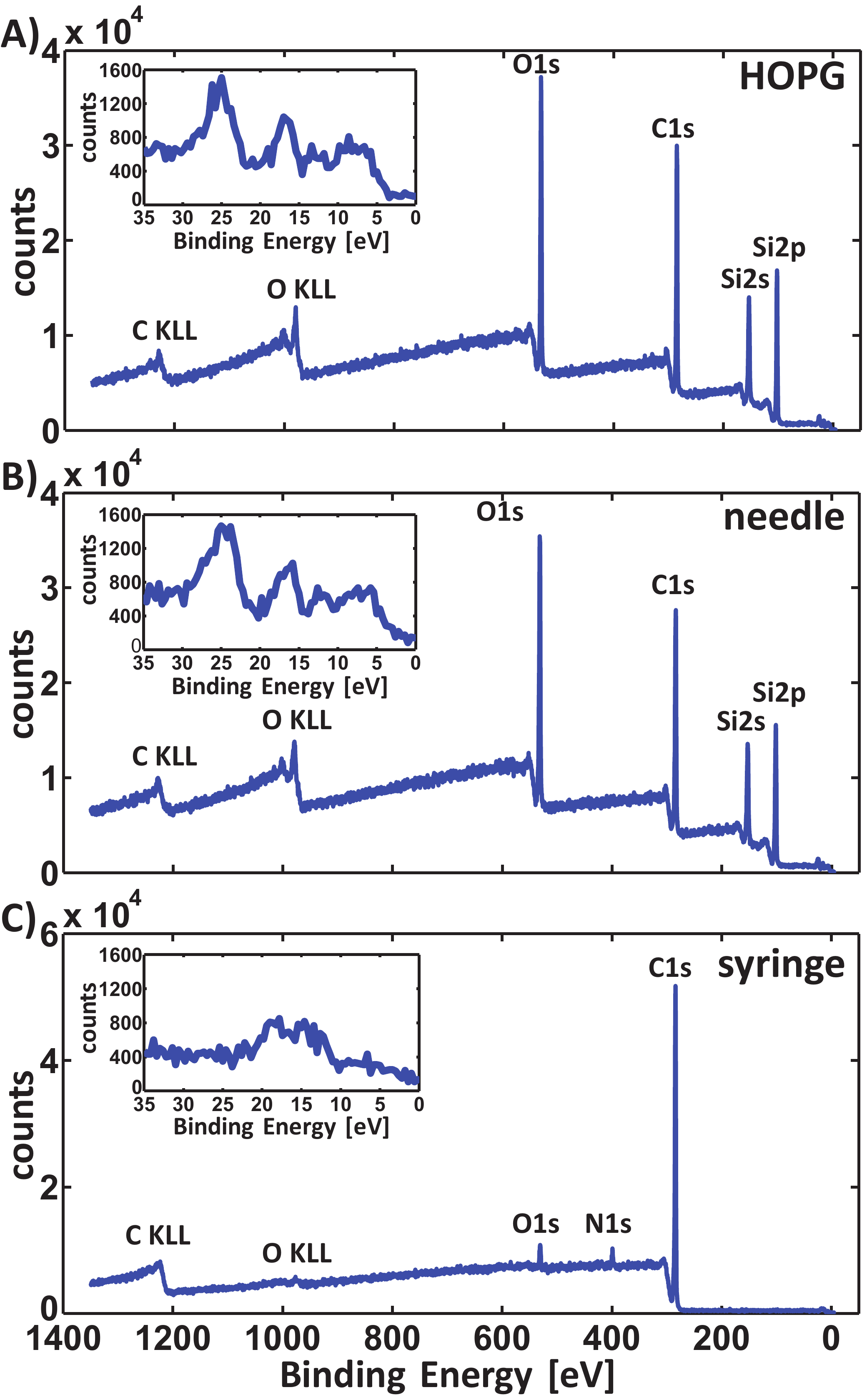}
\caption{ (color online) XPS spectrum of an HOPG sample on which a droplet of water, deposited using a plastic syringe and with a disposable needle, has been dried (A). The peak positions, relative intensity and valence electron spectrum (shown in the inset) indicates that the HOPG surface is covered with a layer of PDMS. The XPS spectrum on the metal cannula of the needle shows a very similar spectrum which can also be attributed to a PDMS layer present (B). The XPS spectrum of the inside of the plastic syringe is completely different and shows no traces of PDMS (C).}
\label{fig:XPS}
\end{figure}

Clearly, in our experiments a contaminant is present in the plastic syringes and/or disposable needles that results in the formation of these features on the surface. The question that remains is: what is the chemical nature of the contamination? To answer this question we performed X-ray photoelectron spectroscopy (XPS) on an HOPG sample on which a droplet of water, deposited using a plastic syringe combined with a disposable needle, was dried. In the resulting XPS spectrum more peaks show up than the normal carbon peak as would be the case for a clean HOPG surface, see Figure~\ref{fig:XPS}A, so there clearly is some contamination on the surface. Table 1 shows a comparison of the atomic percentage, binding energies and O/Si peak ratio of our measurements with XPS data on PDMS from literature. Comparing the peak positions, relative intensity and especially the valence electron spectrum with literature it is possible to chemically characterize the contamination layer as  polydimethylsiloxane (PDMS)~\cite{XPSpolymer}. Since no nanobubble-like objects are observed when using glass syringes with full-metal needles, the contamination has to be induced by the plastic syringe and/or disposable needle. To confirm this, the metal cannula of the disposable needle was measured using XPS, see Figure~\ref{fig:XPS}B. Also in this case a spectrum very similar to that of the dried HOPG was measured, which can be attributed to a $\geq$ 5 nm thick layer of PDMS on the cannula since no metal is visible in the spectrum. Interestingly, XPS measurements on the inside of the plastic syringe do not show any silicon peaks and therefore the syringe itself seems PDMS free, see Figure~\ref{fig:XPS}C. Drying a drop of water, deposited using a plastic syringe and without a disposable needle, on HOPG resulted in a clean XPS spectrum without PDMS contamination.

%
%

%
\begin {table*}
\begin{center}
    \begin{tabular}{ l | c | c | c | c | c | c | c |}
    \cline{2-8}
& \multicolumn{2}{ c| }{\textbf{C1s}} & \multicolumn{2}{ c |}{\textbf{O1s}} & \multicolumn{2}{ c| } {\textbf{Si2p}} & \textbf{O/Si} \\
    \cline{2-8}
 & At. \% & $E_{b}$ [eV] & At. \% & $E_{b}$ [eV] & At. \% & $E_{b}$ [eV] &  \\
\hline
\multicolumn{1}{ |l| }{\textbf{Our data}} &&&&&&& \\
\hline
\multicolumn{1}{ |l| }{Clean HOPG} & 100 & 284.8 &&&&& \\
\hline
\multicolumn{1}{ |l| }{Drying stain on HOPG} & 50.0 & 284.4 & 26.5 & 532.0 & 23.5 & 102.0 & 1.13 \\
\hline
\multicolumn{1}{ |l| }{Disposable needle} & 50.2 & 284.4 & 26.6 & 532.0 & 23.2 & 102.0 & 1.15 \\

\hline \hline
\multicolumn{1}{ |l| }{\textbf{Literature data PDMS}} &&&&&&& \\
\hline
\multicolumn{1}{ |l| }{Beamson and Briggs~\cite{XPSpolymer}} && 284.4 & & 532.00 & & 101.8 &  \\
\hline
\multicolumn{1}{ |l| }{Owen and Smith~\cite{owen1994}} & 50.3 & 285 & 27.1 & & 22.6 & 101.5 & 1.20 \\
    \hline
    \end{tabular}
\end{center}
\caption{ Atomic percentage, binding energy and O/Si peak ratio taken from the XPS measurements on freshly cleaved HOPG,  HOPG on which a droplet of water had dried, deposited using a plastic syringe and disposable needle, and data of an unused disposable needle. Literature data of XPS data on PDMS is shown as a comparison, and shows a clear similarity with our data on the metal needle and contaminated HOPG.}
\label{tab:XPStable}
\end {table*}

%

\begin{figure}[h]
\centering
\includegraphics[width=1\columnwidth]{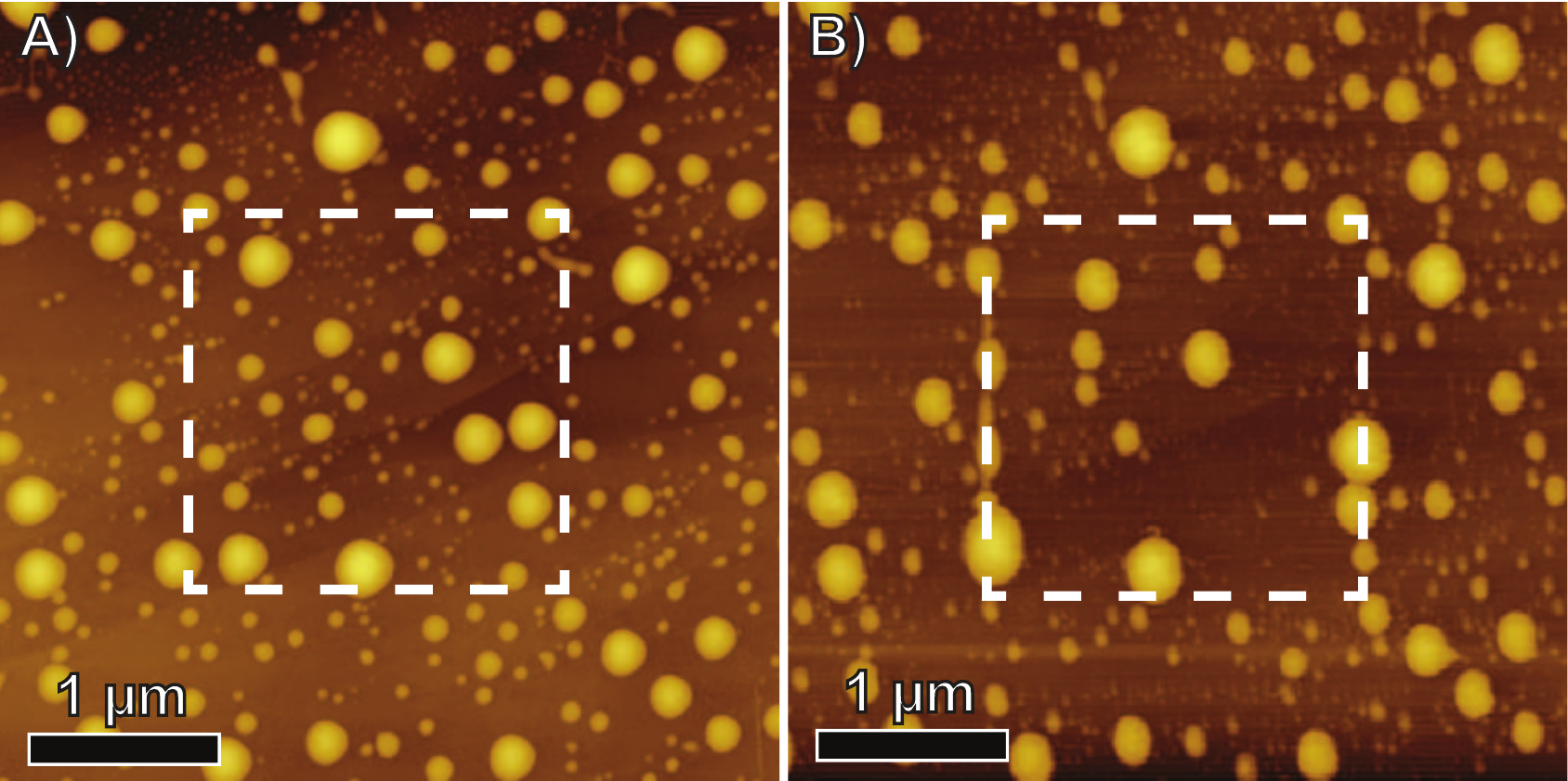}
\caption{ (color online) AFM images of an HOPG surface under a droplet of water mixed with PDMS and deposited using a glass syringe and full-metal needle. Small PDMS droplets have settled on the substrate and have a similar appearance as nanobubbles (A). After scanning a $2 \times 2 ~\unit{\mu m^2}$ area with increased force (highlighted using the dashed square), objects inside this area have moved to another position or coalesced (B). The $z$-range is $21 ~\unit{nm}$. }
\label{fig:coal}
\end{figure}

The formation of nanobubble-like PDMS droplets would be quite consistent with the observation of Evans et al~\cite{evans2004}. In order to confirm whether PDMS contamination is responsible for the nanobubble-like objects we observe, we deliberately added PDMS to our system to confirm the formation of nanobubble-like objects by this polymer. For this we mixed $0.1~\unit{ml}$ of PDMS (Sylgard 184, Dow Corning) with $0.4~\unit{L}$ water by stirring vigorously. A droplet of the PDMS water mixture is then applied to the HOPG substrate using a glass syringe and imaged using the AFM. The resulting AFM images are strikingly similar to that of the nanobubble-like objects produced using plastic syringes with clean Millipore water, see Figure~\ref{fig:coal}A, or from any other nanobubble study for that matter. One of the characteristics of nanobubbles is that it can be moved and coalesced using the AFM tip by increasing the tip-sample interaction. We applied the same technique on the PDMS droplets by scanning a $2 \times 2~\unit{\mu m^2}$ area with increased force (highlighted using a dashed square) and imaging consecutively the same area using normal scanning conditions. Besides the appearance of these PDMS droplet their behavior during increased loads is also strikingly similar to that of nanobubbles reported in literature~\cite{walczyk2013a,walczyk2013,wang2013,zhao2013,borkent2010,yang2008}. The PDMS droplets were moved by the AFM tip and some of the bubbles coalesced, see Figure~\ref{fig:coal}B. The resolution of the AFM image after the scan with increased tip-sample interaction was reduced due to changes of the AFM tip.

%
%

\begin{figure}[h]
\centering
\includegraphics[width=1\columnwidth]{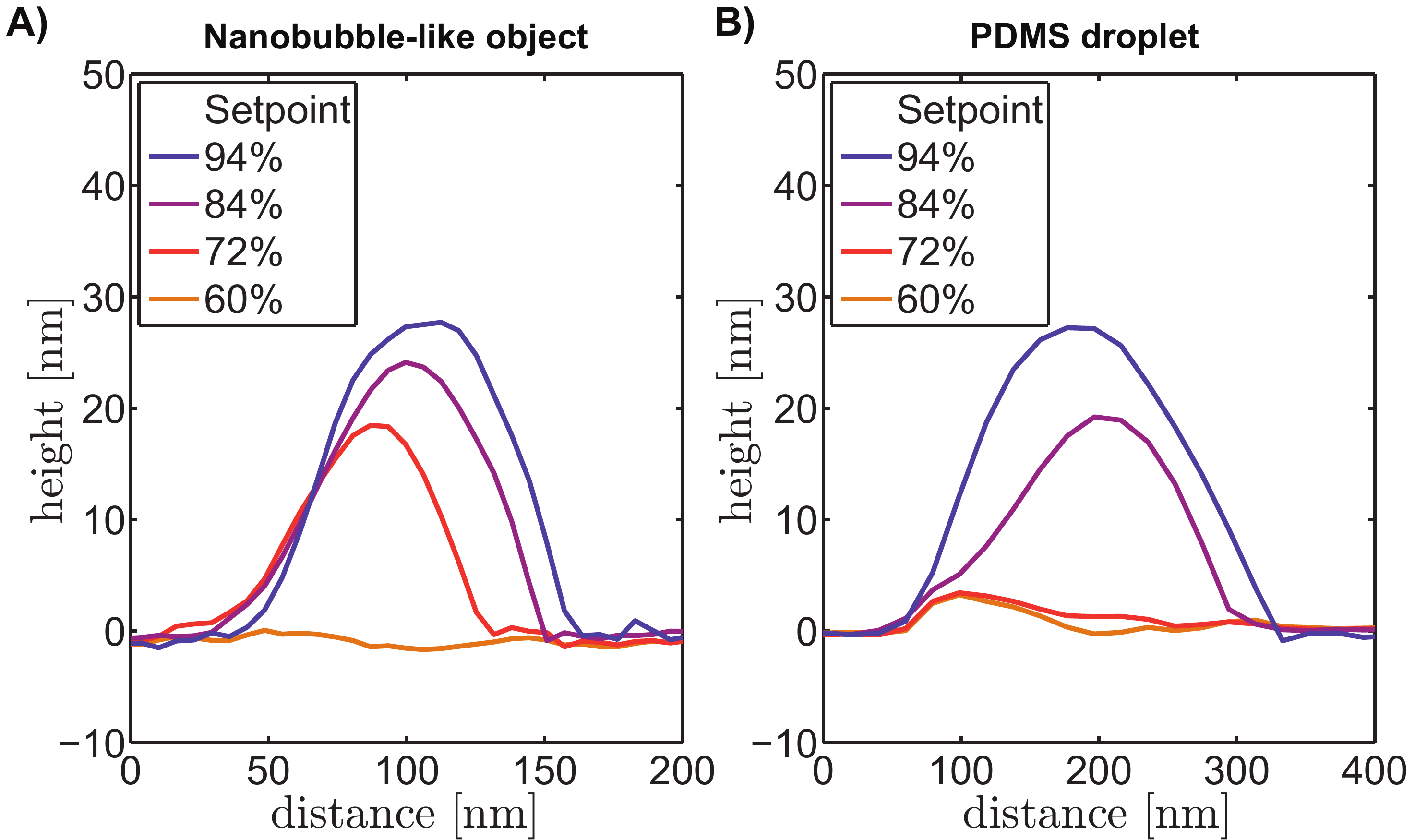}
\caption{ (color online) The dependence of the geometry on four different AFM set-points for a free amplitude of $19~\unit{nm}$ is shown for a nanobubble-like object in water on HOPG produced using a plastic syringe and disposable needle (A). The geometry dependence on four different AFM set-points for a deliberatly added PDMS droplet (using a glass syringe and full-metal needle) on HOPG in water using a free amplitude of $19~\unit{nm}$ shows similar results (B). In both cases the apparent height of the objects is heavily dependent on the set-point.  }
\label{fig:setpoint}
\end{figure}

The literature on nanobubbles states that their apparent shape is very much dependent on the set-point used for scanning the surface in the AFM~\cite{walczyk2013a,walczyk2013,wang2013a,zhao2013,borkent2010}. The set-point dependence of PDMS droplets was therefore compared to that of nanobubble-like objects created using plastic syringes in combination with disposable needles, see Figure~\ref{fig:setpoint}. The radius of curvature, $R_c$, and contact angle, $\theta$,  are $R_c= 90$~nm and $\theta= 50^\circ$ for the object in Figure~\ref{fig:setpoint} A and $R_c= 360$~nm and $\theta= 22^\circ$ for the object in Figure~\ref{fig:setpoint} B, both acquired from the 94 \% set-point measurement.  In both cases the apparent height of the features is heavily dependent on the AFM set-point. A decreasing set-point results in increased tip-sample interactions and soft features like bubbles and droplets are therefore easily deformed by the AFM tip as is the case for the features in these study. 
 
 %
%

The height, radius and contact angle of these features are similar to what is stated in literature on nanobubbles. In addition, these features are soft, could be swept away using an AFM tip and disappeared after drying the surface~\cite{berkelaar2013}. This finding, in retrospect, also may or may not affect some of our and others previous work in which plastic syringes in combination with disposable needles were used\cite{berkelaar2013,berkelaar2012, dietrich2013, borkent2010, borkent2009, borkent2007, walczyk2013, walczyk2013a}. Whether the features observed in these experiments are actual nanobubbles with a gas-impermeable shell induced by the PDMS or are in fact PDMS droplets is something to be investigated and lies outside the scope of this work. The point of the present study is to attain awareness in the nanobubble community, for possible sources of contamination that might in some cases distort experimental results. This could also resolve the mixed results found in literature on a number of subjects related to nanobubbles.

\section{Conclusion}

We have studied the resistance of nanobubble-like objects obtained by droplet deposition on HOPG against a gas-depleted environment using two different experimental techniques. First, the nanobubble-like objects were exposed to a degassed water flow and secondly the ambient pressure was decreased to $20~\unit{mbar}$. In both cases the coverage of the nanobubble-like objects remained virtually unchanged. An in-depth study of possible contamination sources in the procedures and materials used during the experiment showed that in our case the sterile disposable needles were the source of contamination. The chemical nature of the contamination was concluded to be PDMS. Both the nanobubble-like objects and the deliberately formed PDMS droplets can be moved and coalesced using the AFM tip and their apparent shapes depend heavily on the used set-point. The nanobubble-like objects that we nucleated in this way behave not differently from the nanobubbles discussed in literature. The literature on nanobubbles is not in agreement on a variety of subjects. This variance could be resolved by the presence of contamination in some studies, not only originating from disposable needles, influencing the experimental results. We think that it is of utmost importance for the nanobubble community to be aware of the subtlety of contamination sources.

\section{Acknowledgement}
This research was carried out under project number M61.3.10403 in the framework of the Research Program of the Materials innovation institute (M2i, www.m2i.nl). We thank Joost Weijs and Xuehua Zhang for many fruitful discussions.

\begin{appendix}
\section{Appendix A}
Assuming fully gaseous nanobubbles, a reduction of the ambient pressure should lead to an expansion of the bubbles' volume. A sufficient increase in volume could result into contact and coalescence of nanobubbles. The question is: can we expect coalescence for these sizes at such low pressures? To answer this question we calculate the effect of a reduced pressure on nanobubbles, using the assumptions that: (i) there is no mass-transfer between bubble and liquid, (ii) the temperature is constant, and (iii) that either the radius or the contact angle is fixed. Assumption (i) is, in this specific case, justified by the results obtained from the degassed water flow experiment. Assumption (ii) is not completely true since the evaporation of the liquid lowers the temperature, however, this results in only a small change in absolute temperature.  Using the aforementioned assumptions reduces the ideal gas-law to the equation: 

\begin{equation}
P_{1}V_{1}=P_{2}V_{2}.
\label{eq:aa}
\end{equation}

$P$ is the pressure inside the bubble and $V$ the volume of the bubble, where the subscripts 1 and 2 denote the atmospheric and the reduced pressure conditions respectively.
The pressure in the bubble is the result of the combination of ambient pressure and Laplace pressure, $P_{Lap}=2~\gamma / R_{c}$ (the hydrostatic pressure is negligible, for the 5 mm water column it only is $P_{hyd}=\rho g h \approx 0.5~\unit{mbar}$), i.e. eq.~(\ref{eq:aa}) can be rewritten as:

\begin{equation} 
\left(P_{e,1}+\frac{2~\gamma}{R_{c,1}}\right)~V_{1}=\left(P_{e,2}+\frac{2~\gamma}{R_{c,2}}\right)~V_{2}.
\label{eq:ab}
\end{equation}
$P_{e,1}$ and $P_{e,2}$ are the ambient pressures for atmospheric and reduced pressure conditions respectively. $R_{c,1}$ and $R_{c,2}$ are the radii of curvature of the nanobubble in atmospheric and reduced pressure conditions, and $\gamma$ is the surface tension of water, $\gamma=72~\unit{mN/m}$.
Using the spherical cap geometry of a nanobubble to acquire a relation for the nanobubble volume results in the following equations:

\begin{equation} 
V(R_{c},h)=\frac{1}{3}~\pi~h^{2}~(3~R_{c}-h),
\label{eq:ae}
\end{equation}

\begin{equation}
R_{c}(r,\theta)=\frac{r}{\sin \theta},
\label{eq:af}
\end{equation}

\begin{equation} 
h(r,\theta)=\frac{r}{\sin \theta}-r \cot \theta,
\label{eq:ah}
\end{equation}
where $h$ is the height of the nanobubble, $r$ is the radius of the contact line, and $\theta$ is the contact angle (measured in the gas-phase).
Combination of eq.~(\ref{eq:ae}) to (\ref{eq:ah}) results in the following expression for the nanobubble volume:

\begin{equation} 
V(r,\theta)=X(\theta)~r^3,
\label{eq:ai}
\end{equation}

\begin{equation} 
X(\theta)=\frac{1}{3}~\pi  \left(\frac{2}{\sin \theta}+\cot \theta \right) \left(\frac{1}{\sin \theta}-\cot \theta \right)^{2}.
\label{eq:ad}
\end{equation}
So with relations eq.~(\ref{eq:af}),~(\ref{eq:ai}), and~(\ref{eq:ad}), eq.~(\ref{eq:ab}) is transformed into a function of $r$ and $\theta$:

\begin{equation} 
\left(P_{e,1}+\frac{2~\gamma~\sin \theta_{1}}{r_{1}}\right)~X(\theta_{1})~r_{1}^{3}=\left(P_{e,2}+\frac{2~\gamma~\sin \theta_{2}}{r_{2}}\right)~X(\theta_{2})~r_{2}^{3}.
\label{eq:ac}
\end{equation}

%
\begin{figure}[h]
\centering
\includegraphics[width=1\columnwidth]{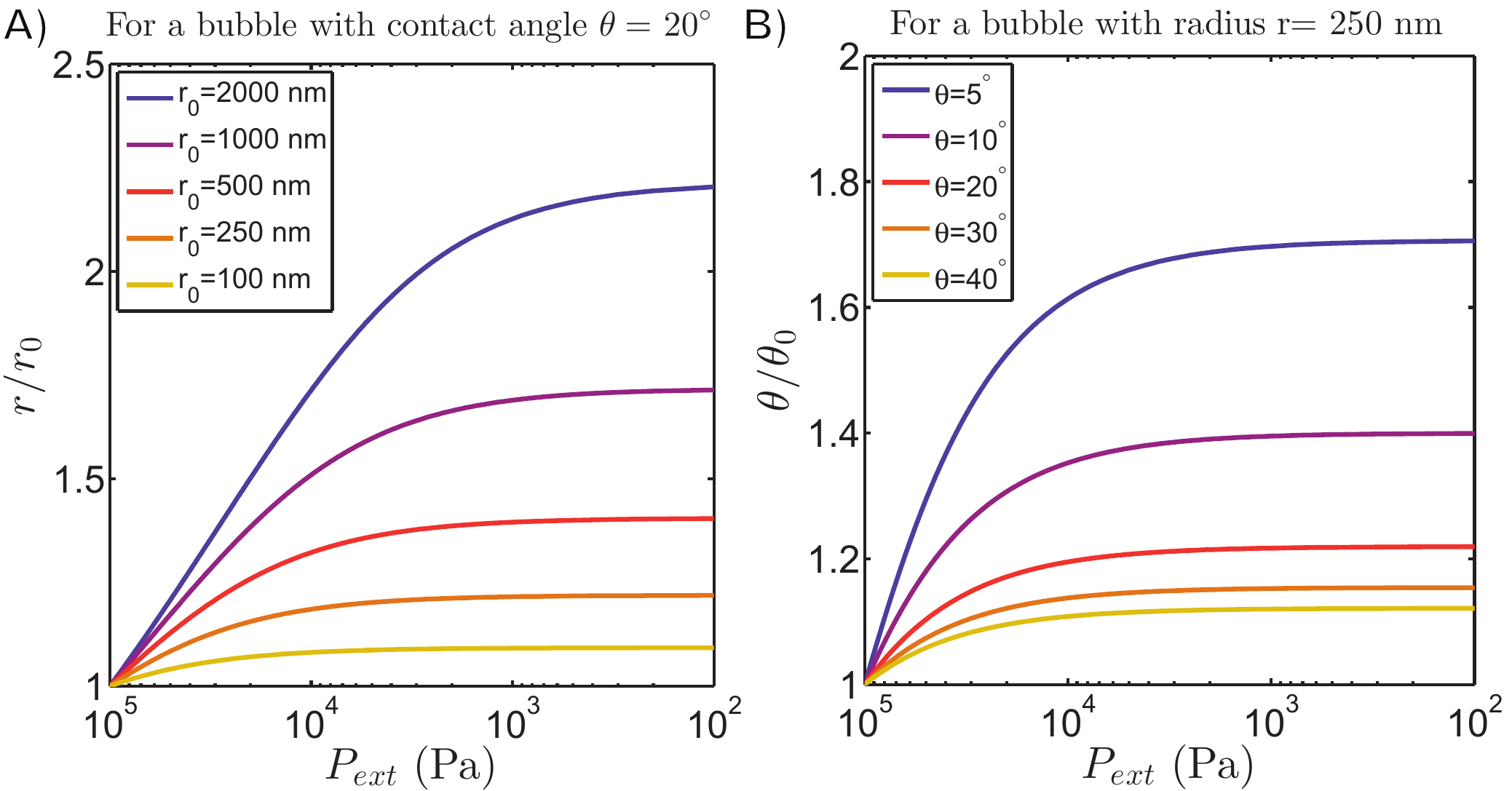}
\caption{(Color online) The radius of a nanobubble versus the ambient pressure (note the inversion of the axes) for a fixed contact angle of $20^{\circ}$ is calculated numerically using eq.~(\ref{eq:ac}) (A). The increase in radius is limited for bubbles with a radius of $1000~\unit{nm}$ and less. In the case of a pinned contact line (fixed radius at $r=250~\unit{nm}$, i.e. fixed footprint area) the contact angle increases only by a fraction of its original value (B). The contact angle never exceeds $90^{\circ}$. }
\label{fig:theory}
\end{figure}
%

Assuming a \emph{fixed contact angle} of $20^{\circ}$ ($160^{\circ}$ in the liquid phase), which is a typical value for nanobubbles~\cite{borkent2010}, we numerically calculate the change in radius as a function of pressure from $1~\unit{atm}$ to $1~\unit{mbar}$ (100 Pa) for different initial radii, see Figure~\ref{fig:theory}A. Remarkably, for typical nanobubbles, which have a radius between 100 and $1000~\unit{nm}$, the increase in radius is rather limited. For the second case of a \emph{fixed radius} of the nanobubble (fixed at $250~\unit{nm}$),  the expansion in volume is achieved by an increase in contact angle, see Figure~\ref{fig:theory}B. Also in this case the contact angle increases only with a fraction of the original value, always remaining smaller than $90^{\circ}$, so the lateral size remains constant. The limited increase in contact angle or radius can be explained by the enormous Laplace pressure for small bubbles; a reduction in ambient pressure leads to only a small reduction in the bubbles' internal pressure. Therefore, it is not surprising that no coalescence of surface nanobubbles is observed at a reduced pressure of $20~\unit{mbar}$.

\end{appendix}



\footnotesize{
\bibliography{references} 
\bibliographystyle{rsc} 
}

\end{document}